\documentclass[12pt]{iopart}

\usepackage{graphicx}

\begin{document}

\title[Is the Fair Sampling Assumption supported by EPR
Experiments?]{Is the Fair Sampling Assumption supported by EPR
Experiments?}

\author{Guillaume Adenier, Andrei Yu. Khrennikov}

\address{International Center for Mathematical Modelling in
Physics and Cognitive Sciences\\V\"axj\"o University, 351 95
V\"axj\"o, Sweden.}

\ead{guillaume.adenier@vxu.se}

\begin{abstract}
We analyze optical EPR experimental data performed by Weihs \emph{et
al.} in Innsbruck 1997-1998. We show that for some linear
combinations of the raw coincidence rates, the experimental results
display some anomalous behavior that a more general source state
(like non-maximally entangled state) cannot straightforwardly
account for. We attempt to explain these anomalies by taking account
of the relative efficiencies of the four channels. For this purpose,
we use the fair sampling assumption, and assume explicitly that the
detection efficiencies for the pairs of entangled photons can be
written as a product of the two corresponding detection efficiencies
for the single photons. We show that this explicit use of fair
sampling cannot be maintained to be a reasonable assumption as it
leads to an apparent violation of the no-signalling principle.
\end{abstract}

\pacs{03.65.-w, 03.65.Ud}

\maketitle

\section{Introduction}

The experimental violation of Bell Inequalities
\cite{epr,bohm,bell64,CHSH} in optical EPR experiments
\cite{Aspect82,Weihs} can only be validated under the additional
assumption of \emph{fair sampling} \cite{CHSH,ClauserS,AdenKhren02}.
This type of test being crucial for modern quantum communication
research \cite{Tittel,Jaeger}, it is unfortunate that the result of
a test meant to disprove Local Realism would depend on such an
additional assumption, since Local Realism is a priori no less
plausible an assumption than Fair Sampling is.

In this article we examine data from past experiments and look for
possible traces of failure of the Fair Sampling assumption. For this
purpose, we were recently given the chance to examine raw data from
the long distance experiment with fast switching performed in
Innsbruck in 1997-1998 by Gregor Weihs \emph{et al}. We show that
the near perfect agreement with the predictions of Quantum Mechanics
obtained from experiments can be questioned when extracted data are
looked from a different perspective, that is, when comparing other
linear combinations of the normalized coincidence rates than the
correlation function with the predictions of Quantum Mechanics.

\section{Channel efficiencies}

In a real experiment, not all photons are detected and one should
take account of the efficiencies of each of the four channels
involved. This problem is at the heart of the detection efficiency
loophole, and can be used to design local realistic models that
reproduce the main experimental features of optical EPR experiments
\cite{Larsson}.

We label $\eta_A^{+}(\alpha)$ and $\eta_A^{-}(\alpha)$ the
single-count channel efficiencies of Alice's plus and minus
channels respectively, and similarly $\eta_B^{+}(\beta)$ and
$\eta_B^{-}(\beta)$ for Bob's plus and minus channels; the
parameter dependence in $\alpha$ and $\beta$ reflects the fact
that these efficiencies may, in principle, depend on the local
settings. It is important to stress that we can put in the concept
of \emph{channel efficiencies} all possible variations in the
experimental conditions that are not directly related to the
quantum state associated to the source (such as signal intensity
variations, detector efficiencies, optical misalignments,
collection efficiencies, etc.). This idea is important since it is
precisely the aim of our paper to analyze experimental quantities
that can directly be compared with Quantum predictions --- under
the assumption of Fair Sampling --- independently of all these
experimental contingencies encapsulated in the channel
efficiencies.

We can expect the number of single counts to be proportional to
the respective channel efficiencies:
\begin{eqnarray}\label{singlec}
    \nonumber
    N^{A,\varepsilon_1}_{\rm exp}(\alpha)\approx\eta_A^{\varepsilon_1}(\alpha)N_{A}/2\\
    \nonumber
    N^{B,\varepsilon_2}_{\rm
    exp}(\beta)\approx\eta_B^{\varepsilon_2}(\beta)N_{B}/2,
\end{eqnarray}
where $N_{A}$ and $N_{B}$ are the (unknown) number of photons
actually sent respectively to Alice and Bob, and where
$\varepsilon_1$ and $\varepsilon_2$ can each be either $+$ or $-$,
to shorten the forthcoming equations\footnote{The use of the symbol
$\approx$ is here to remind of the statistical variability in any
experimental results that is naturally expected to induce small
deviation from the predictions. The amplitude of these deviations
are expected to decrease with the number of trials obtained
experimentally.}.

We now consider the number of coincidence counts experimentally
registered. If we consider a pair of photons as a whole, the
probability that it is detected in each specific combination of
two channels should depend as well on a combined channel
efficiency. We label these combined channel efficiencies as
$\eta^{++}(\alpha,\beta)$, $\eta^{+-}(\alpha,\beta)$,
$\eta^{-+}(\alpha,\beta)$, and $\eta^{--}(\alpha,\beta)$. The
number of coincidences in a pair of channels
$(\varepsilon_1,\varepsilon_2)$ should thus be proportional to the
relevant combined efficiency
$\eta^{\varepsilon_1,\varepsilon_2}(\alpha,\beta)$, the (unknown)
number of pairs sent $N_{AB}$, and the relevant joint probability
predicted by Quantum Mechanics
$P_{QT}^{\varepsilon_1,\varepsilon_2}(\alpha,\beta)$, that is,
\begin{equation}
    N^{\varepsilon_1,\varepsilon_2}_{\rm
    exp}(\alpha,\beta)\approx\eta^{\varepsilon_1,\varepsilon_2}(\alpha,\beta)\;N_{AB}\;P_{QT}^{\varepsilon_1,\varepsilon_2}(\alpha,\beta).\\
\end{equation}
Here the angular dependence of the efficiencies is clearly unwanted
since we are interested in an experimental test of the predictions
of Quantum Mechanics, independently of the inaccuracies involved in
any particular experimental setup. Our immediate purpose is
therefore to get rid of these angular dependencies due to the
combined efficiencies.

In order to do so, we assume that the ensemble of detected pairs
of photon provides a fair statistical sample of the ensemble of
emitted pairs (Fair Sampling Assumption). A consequence of this
assumption is that the probabilities of non detection for Alice
and Bob should be independent of one another. Indeed, for fixed
settings $(\alpha,\beta)$, the probability of a non detection
should be independent of the polarization state of the photon
(otherwise the sampling would clearly be unfair), and should thus
be independent of the fate of the distant correlated photon.

Hence, the channel efficiencies should be the same for all photons
going into a specific channel, independently of whether a photon
happens to be single or to be paired with a distant detected photon.
That is, the above combined efficiencies for pairs of particles
should be equal to the product of the relevant channel efficiencies
for the single counts:
\begin{equation}\label{FSassumption}
    \eta^{\varepsilon_1,\varepsilon_2}(\alpha,\beta)=\eta_A^{\varepsilon_1}(\alpha)\eta_B^{\varepsilon_2}(\beta).
\end{equation}

With this factorization of the channel efficiencies, we can rewrite
the predicted number of coincidence counts as:
\begin{equation}\label{Ncoinc}
    N^{\varepsilon_1,\varepsilon_2}_{\rm exp}(\alpha,\beta)\approx N_{AB}\;\eta_A^{\varepsilon_1}(\alpha)\eta_B^{\varepsilon_2}(\beta)\;P_{QT}^{\varepsilon_1,\varepsilon_2}(\alpha,\beta)\\
\end{equation}

\section{Normalizing using the single counts}

The probabilities derived from the standard normalization
procedure with coincidence counts still depend explicitly on the
channel efficiencies, so that some properties of the photon pairs
will remain out of reach of the experimenter.

True enough, in some specific cases the standard normalization
procedure can make the channel efficiencies disappear from the
final result. For instance, if on one side the channels are
balanced, (e.g. $\eta_B^{+}(\beta)=\eta_B^{-}(\beta)$), it is
straightforward to show that the standard normalization with the
total sum of coincidences removes the channel efficiency
dependence of the correlation function:
\begin{eqnarray}\label{etacorr}
    \frac{N^{++}_{\rm exp}+N^{--}_{\rm exp}-N^{+-}_{\rm
    exp}-N^{-+}_{\rm exp}}{\sum_{\varepsilon_1,\varepsilon_2} N^{\varepsilon_1,\varepsilon_2}_{\rm exp}}\approx
    P^{++}_{QT}+P^{--}_{QT}-P^{+-}_{QT}-P^{-+}_{QT}=E_{QT}.
\end{eqnarray}

However, for other linear combination of the coincidence counts,
the normalization by the total sum of coincidences cannot remove
the channel efficiency dependence, even if on one side the
channels are balanced. For instance, if
$\eta_B^{+}(\beta)=\eta_B^{-}(\beta)$, then:
\begin{eqnarray}\label{etamarg}
    \frac{N^{++}_{\rm exp}+N^{+-}_{\rm
    exp}}{\sum_{\varepsilon_1,\varepsilon_2} N^{\varepsilon_1,\varepsilon_2}_{\rm exp}}\approx \frac{\eta_A^{+}}{\eta_A^{+}+\eta_A^{-}}
    (P^{++}_{QT}+P^{+-}_{QT})\neq(P^{++}_{QT}+P^{+-}_{QT}).
\end{eqnarray}

This problem makes it necessary to circumvents these dependencies
in all cases by means of the new normalization procedure. Our idea
is that most of the counts registered in the channels are
non-coincidence events, because the channel efficiencies are low.
Since there are many times more non-coincident events than
coincident ones, they provide a useful and accurate additional
statistical information about the relative efficiency of the
channels.

In order to get rid of channel efficiencies in the above
equations, we thus define the following experimental quantities,
which are proportional to the ratio of the number of coincidence
counts in a combined channel over the product of the two
corresponding single counts:
\begin{equation}\label{defratio}
    f_{\rm exp}^{\varepsilon_1,\varepsilon_2}(\alpha,\beta)\equiv\frac{1}{4}\frac{N_{\rm exp}^{\varepsilon_1,\varepsilon_2}(\alpha,\beta)}{N^{A,\varepsilon_1}_{\rm exp}(\alpha)N^{B,\varepsilon_2}_{\rm exp}(\beta)}
\end{equation}
Replacing the single counts and coincidence counts by their
expressions given respectively in (\ref{singlec}) and
(\ref{Ncoinc}), we obtain:
\begin{equation}
    f_{\rm exp}^{\varepsilon_1,\varepsilon_2}(\alpha,\beta)\approx\frac{N_{AB}}{N_{A}N_{B}}\;P_{QT}^{\varepsilon_1,\varepsilon_2}(\alpha,\beta)\\
\end{equation}
for which the only angular dependence is the one due to the Quantum
Mechanical term.

Since the four joint probabilities $P_{QT}^{++}(\alpha,\beta)$,
$P_{QT}^{+-}(\alpha,\beta)$, $P_{QT}^{-+}(\alpha,\beta)$ and
$P_{QT}^{--}(\alpha,\beta)$ add up to unity, summing these four
equation together yields
\begin{equation}
\sum_{\varepsilon_1,\varepsilon_2} f_{\rm
exp}^{\varepsilon_1,\varepsilon_2}(\alpha,\beta)\approx\frac{N_{AB}}{N_{A}N_{B}}.
\end{equation}
and finally,
\begin{eqnarray}\label{Qcomp}
P_{QT}^{\varepsilon_1,\varepsilon_2}(\alpha,\beta)\approx\frac{f_{\rm
exp}^{\varepsilon_1,\varepsilon_2}(\alpha,\beta)}{\sum_{\varepsilon_1,\varepsilon_2}
f_{\rm exp}^{\varepsilon_1,\varepsilon_2}(\alpha,\beta)}
\end{eqnarray}

Since the normalization procedure that we propose here is different
than the standard one based on the coincidence counts alone, a word
on the relevance of Eq.(\ref{Qcomp}) with experiments might be
necessary here. The validity of Eq.(\ref{Qcomp}) as an approximation
of quantum probabilities by experimental frequencies (based on the
single counts) depends only on the validity of the Fair Sampling
Assumption, an assumption routinely made in EPR experiments
exhibiting a violation of Bell inequalities\footnote{To some extent,
one could argue that John Bell used an implicit Fair Sampling
Assumption in deriving his theorem, and that although he discarded
the possibility that it would not be fulfilled as very unlikely, it
remained however a necessary condition that had to always be
assumed, but was never fully tested experimentally. In that sense,
the Fair Sampling Assumption is as essential to the whole framework
of Bell's theorem as are the Locality assumption and the Realism
assumption.}. Indeed, Eq. (\ref{FSassumption}) is but a consequence
of this assumption, and leads straightforwardly to Eqs.
(\ref{defratio}), (\ref{Qcomp}), and (\ref{probas}).

Hence, by using the Fair Sampling Assumption together with
experimental statistics for the single counts, we are able to
obtain experimental quantities that should coincide with the
prediction given by Quantum Mechanics for the joint probabilities,
independently of any channel efficiency imbalance. As far as we
know, this procedure is new and offers new perspectives for
comparing experimental result with Quantum predictions.

In particular this new normalization procedure allows us to obtain
experimental quantities that should directly coincide with the
marginal probabilities:
\begin{eqnarray}\label{probas}\nonumber
\frac{f_{\rm exp}^{++}(\alpha,\beta)+f_{\rm
exp}^{+-}(\alpha,\beta)}{\sum_{\varepsilon_1,\varepsilon_2} f_{\rm
exp}^{\varepsilon_1,\varepsilon_2}(\alpha,\beta)}&\approx&P^{++}_{QT}(\alpha,\beta)+P^{+-}_{QT}(\alpha,\beta)
\\
\frac{f_{\rm exp}^{-+}(\alpha,\beta)+f_{\rm
exp}^{--}(\alpha,\beta)}{\sum_{\varepsilon_1,\varepsilon_2} f_{\rm
exp}^{\varepsilon_1,\varepsilon_2}(\alpha,\beta)}&\approx&P^{-+}_{QT}(\alpha,\beta)+P^{--}_{QT}(\alpha,\beta)
\\
\frac{f_{\rm exp}^{++}(\alpha,\beta)+f_{\rm
exp}^{-+}(\alpha,\beta)}{\sum_{\varepsilon_1,\varepsilon_2} f_{\rm
exp}^{\varepsilon_1,\varepsilon_2}(\alpha,\beta)}&\approx&P^{++}_{QT}(\alpha,\beta)+P^{-+}_{QT}(\alpha,\beta)
\nonumber
\\
\frac{f_{\rm exp}^{+-}(\alpha,\beta)+f_{\rm
exp}^{--}(\alpha,\beta)}{\sum_{\varepsilon_1,\varepsilon_2} f_{\rm
exp}^{\varepsilon_1,\varepsilon_2}(\alpha,\beta)}&\approx&P^{+-}_{QT}(\alpha,\beta)+P^{--}_{QT}(\alpha,\beta).
\nonumber
\end{eqnarray}

An important prediction of Quantum Mechanics for these marginal
probabilities is the no-signalling principle. A correlation
$P(\varepsilon_1,\varepsilon_2|\alpha,\beta)$ is non-signaling if
and only if its marginal probabilities are independent of the
other side input: $\sum_{\varepsilon_2}
P(\varepsilon_1,\varepsilon_2|\alpha,\beta)$ is independent of
$\beta$ and $\sum_{\varepsilon_1}
P(\varepsilon_1,\varepsilon_2|\alpha,\beta)$ is independent of
$\alpha$ \cite{Gisin}.

This non-signalling property does not depend on the state of the
source of photons. Whatever is the source sent to Alice and Bob,
they cannot use it to communicate. Alice's marginal probabilities,
represented by the two first equations above, cannot depend on
Bob's measurement setting $\beta$. Similarly, Bob's marginal
probabilities, represented by the two last equations above, cannot
depend on Alice's measurement setting $\alpha$.

Hence, if in a series a measurement only one of these parameter
varies (say, $\alpha$), then only two of these four quantities can
vary accordingly, the other two remaining constant. It is
precisely this no-signaling prediction conditioned on the validity
of the Fair Sampling Assumption that we want to check here.

\section{Experimental Results}
\subsection{Long distance with fast switching}

Out of all the various runs we had at hand, we chose first a set of
files containing data from a long distance fast switching
experiment, with a significant number of coincidences and with
enough measurement angles. This run was performed on the 1st of may
1998 in Innsbruck. The files were referred to as \emph{Scanblue},
meaning that a scan varying "blue" (for Alice) side modulator bias
was performed, with both sides randomly fast switching between an
equivalent +0 and +45 degrees angle. Modulating the bias from -100
to +100 was linearly equivalent to rotating the corresponding
Polarizing Beam splitter from $-\frac{\pi}{2}$ to $+\frac{\pi}{2}$.

\begin{figure}
\includegraphics[width=8cm]{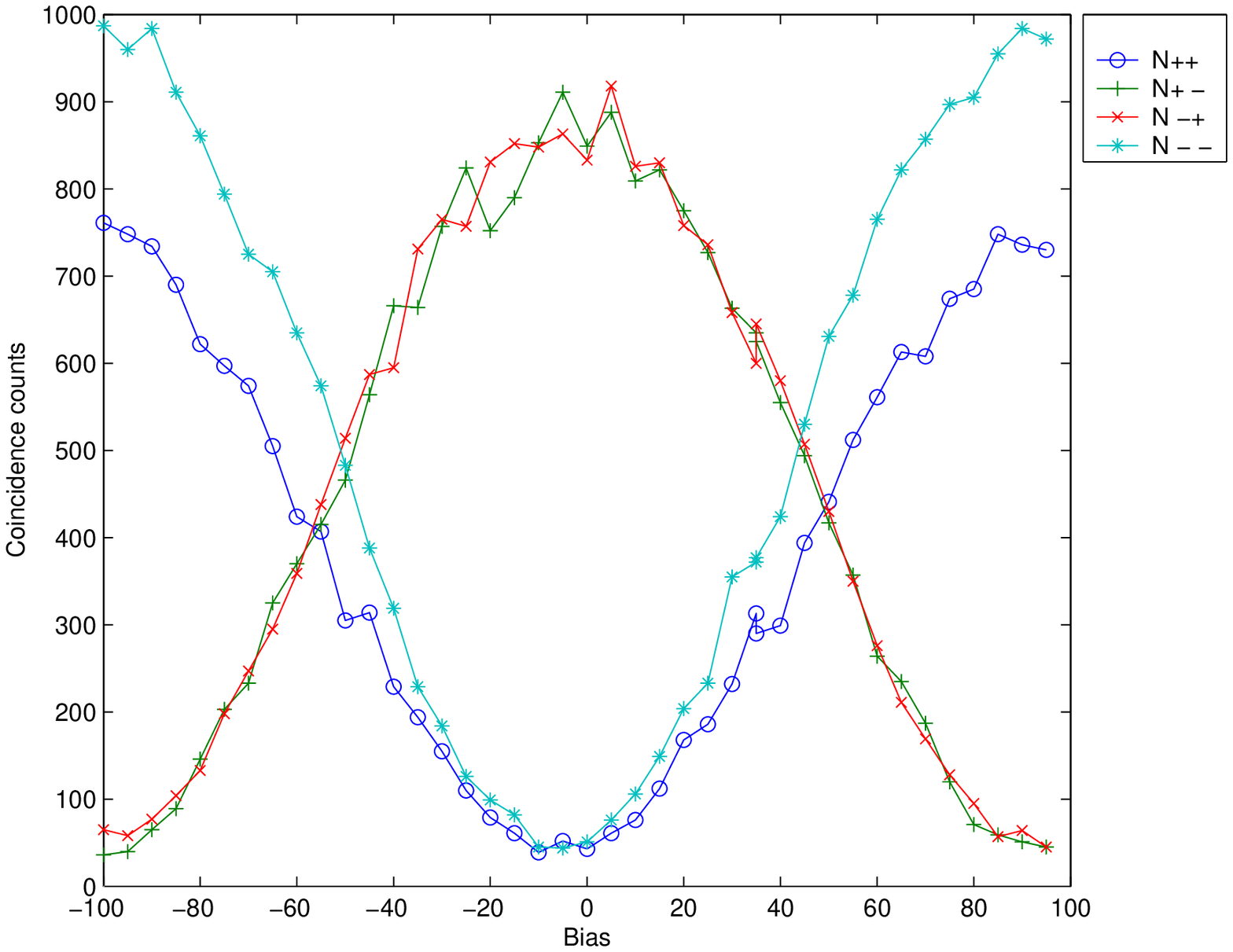}
\includegraphics[width=8cm]{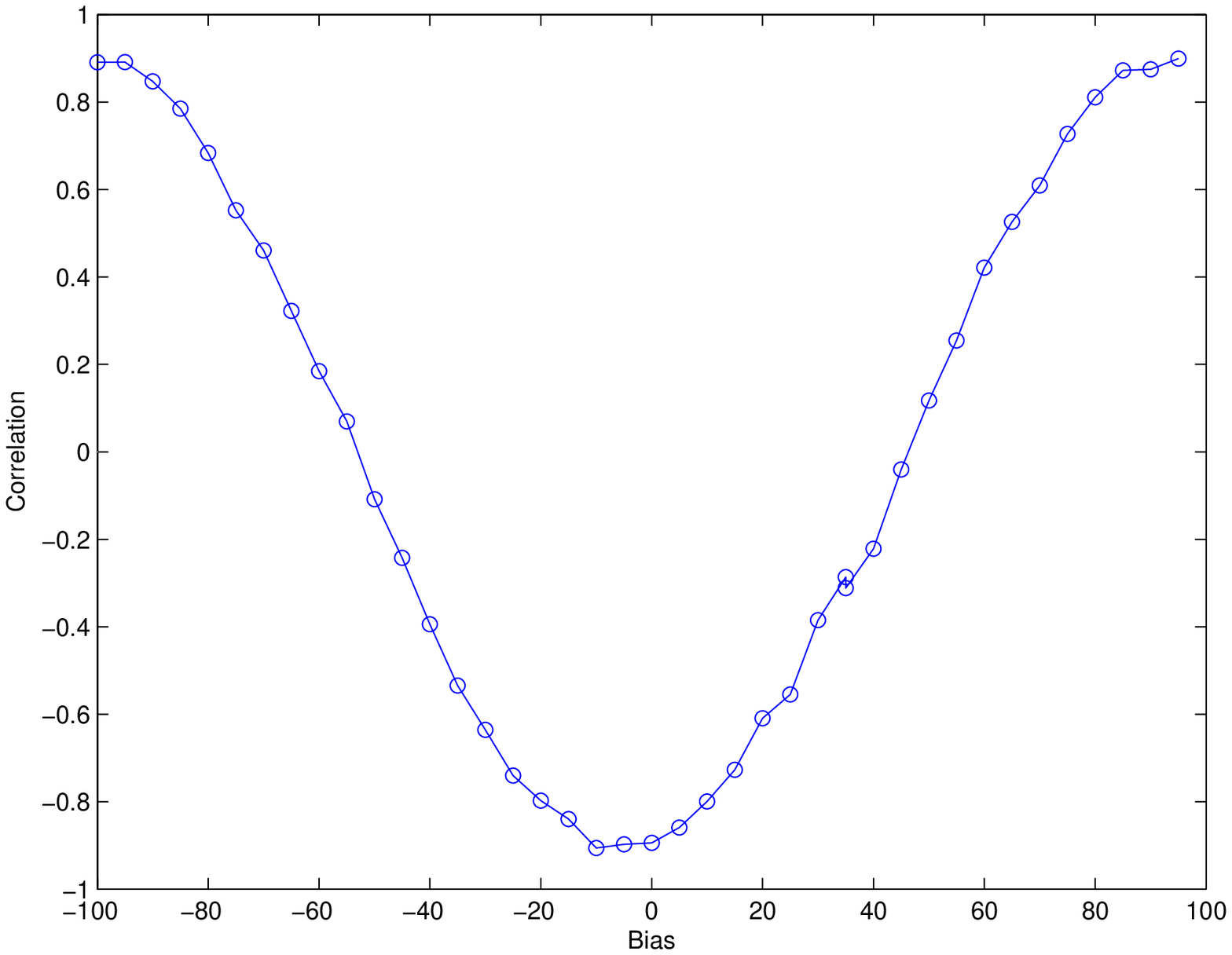}
\caption{\label{fig:rawfig1} {\bf (a)} Coincidence counts for a long
distance Bell test with fast measurement switching (Innsbruck 1998,
source file: \emph{scanblue}, with switches on position 00),
\\{\bf (b)} Correlation Function.}
\end{figure}

We look at the data for the particular case when both switches were
set to +0 degrees angle. As can be seen in the coincidence rate
figures (see figure \ref{fig:rawfig1}a), the coincidence rates
exhibit minima close to zero, and cosine-squared shape, as expected
from the predictions of Quantum Mechanics. However, the maxima of
the four coincidence curves differ significantly.
In spite of this anomalous behavior, the correlation function
computed with the standard normalization (i.e., with the sum of all
coincidence counts) coincides very well with the quantum mechanical
predictions (see figure \ref{fig:rawfig1}b).

It is interesting to note that possibly similar anomalies were
observed first in the two-channel EPR experiments performed by
Alain Aspect in Orsay in the early 1980's. The anomalies were
reported in Aspect's PhD Thesis \cite{AspectTh} in the following
words:
\begin{quote}
\emph{For some measurements, we have observed abnormal differences
between coincidence rates that were expected to be equal (for
instance $N_{+-}$ and $N^{-+}$). It turns out however that even for
these measurements the correlation coefficient $E(a,b)$ remains
equal to the quantum predictions, with better than two standard
deviations. We have no completely convincing explanation, either for
these anomalies, or for their compensation} \footnote{Our
translation. The original text goes as follow: \emph{Pour quelques
mesures on a observ\'{e} des diff\'{e}rences anormales entre taux de
coincidences que l'on attendait \'{e}gaux (par exemple $N_{+-}$ et
$N_{-+}$). Mais il se trouve que pour ces mesures, le coefficient de
corr\'{e}lation $E(a,b)$ reste \'{e}gal à la pr\'{e}vision
quantique, \`{a} mieux que deux \'{e}carts-types pr\`{e}s. Nous
n'avons aucune explication compl\`{e}tement convaincante, ni pour
ces anomalies, ni pour leur compensation.}}.
\end{quote}
This anomalies could in principle be explained by a non-rotationally
invariant source state, such as a non-maximally entangled state (see
\ref{QMIdeal}). The information available on these anomalies being
reduced to the above quotation, it is however impossible to give a
definite explanation for them. These anomalies are nevertheless
interesting since they indicate that, assuming the state produced by
the source was a singlet state and that the channel efficiencies
where balanced on each side, the experimental data would not have
coincided with the predictions of Quantum Mechanics for all possible
linear combinations of the normalized coincidence counts, although
the fit with the correlation function was excellent.

\begin{figure}
\center
\includegraphics[width=8cm]{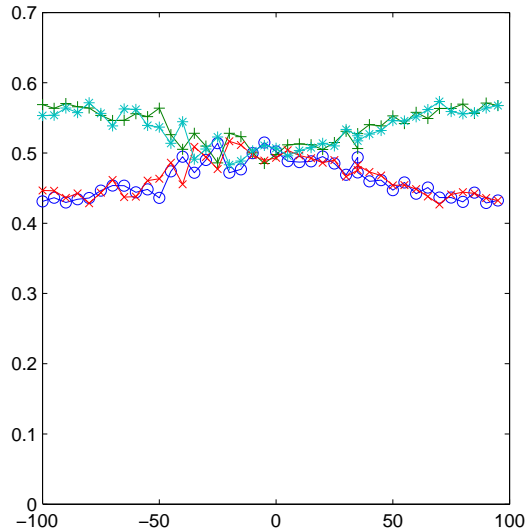}
\caption{\label{fig:NoddplusNeven}The four possible curves with
$(N^{even}_{\rm exp}+N^{odd}_{\rm
    exp})/ \sum_{\varepsilon_1,\varepsilon_2} N^{\varepsilon_1,\varepsilon_2}_{\rm
    exp}$, showing that the standard normalization with coincidence counts leads to
incorrect results.}
\end{figure}

\begin{figure}
\center
\includegraphics[width=8cm]{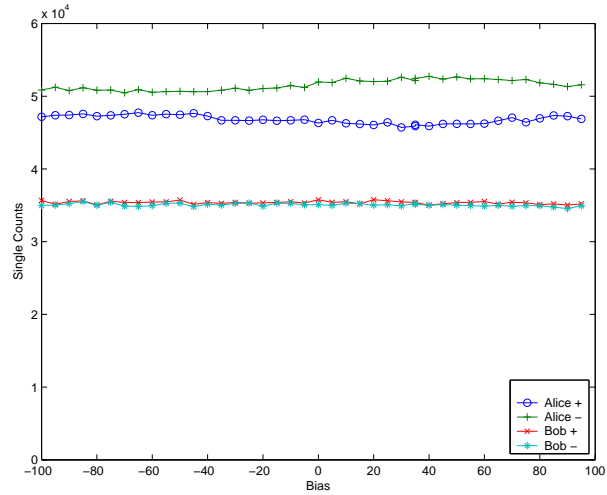}
\caption{\label{fig:singlesfig} Single Counts (Switches 00,
scanblue).}
\end{figure}

To illustrate this point with our particular set of data, we
computed the four possible sum of \emph{even} normalized coincidence
counts with \emph{odd} normalized coincidence counts (see figure
\ref{fig:NoddplusNeven}). In an ideal setup with balanced channels,
these four curve should coincide with Alice's and Bob's marginal
probabilities (that is, $1/2$ for a singlet state).
The result displayed in figure \ref{fig:NoddplusNeven} shows
clearly that other linear combination of the normalized
coincidence counts than the correlation function leads to
incorrect results. For this purpose we extract first the single
counts. The result is displayed in figure \ref{fig:singlesfig}.

As can be seen from the figure, some slight variations of Alice's
single count rates can be observed as she varies her bias. It is
nevertheless of small amplitude, and remains in any case always
local, in the sense that only Alice's single counts vary. More
important is the fact that the efficiency of + and - channels differ
significantly for Alice.

\begin{figure}
\center
\includegraphics[width=8cm]{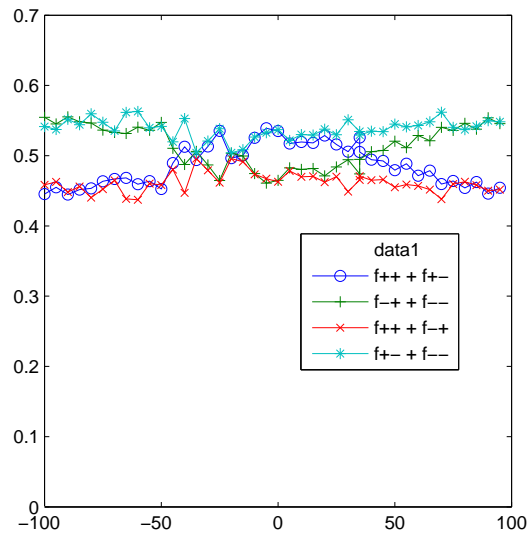}
\caption{\label{fig:marg1} These four quantities
$(f^{even}+f^{odd})/\sum f$ should coincide with Alice's and Bob's
marginal probabilities.}
\end{figure}

As can be seen from figure \ref{fig:marg1}, the \emph{four}
experimental quantities expressed by (\ref{probas}) vary with
Alice's measurement angle $\alpha$.

For the two first, corresponding to Alice's marginal probabilities,
this dependence can be accounted for by a non-rotationally invariant
source state, such as a non-maximally entangled state (see
Appendix). However, for the two remaining, there should be no angle
dependence on $\alpha$ at all, since they should depend only on
$\beta$ which remains fixed. The angular dependence of the
experimental quantities that should coincide with Bob's marginal
probabilities is admittedly weak, but statistical analysis show that
a linear fit provides a much poorer fit than a non linear fit for
these plots.

Nevertheless, since we had other sets of data available, we decided
to investigate whether this anomalous behavior could be exhibited in
other experimental setup that would not necessarily be meant to
close the locality loophole, but would have cleaner results.

\subsection{Short distance without fast switching}

We thus chose a set of data that have a relatively high maximum
number of coincidences and low dark rates, as well as enough
different measurement settings, in order to exhibit better the
rather tenuous anomaly we observed in all the runs. This run was
performed 1997 in Innsbruck. The files were referred to as
\emph{Bluesine}, meaning that only the "blue" side (that is Alice)
varied the measurement setting $\alpha$, while the "red" side (that
is, Bob) kept the same setting. It had no fast switching and was
performed at short distance, but since our goal was not to focus on
the locality loophole but rather on detection and sampling issues,
this would have been irrelevant anyway.

\begin{figure}
\center
\includegraphics[width=7cm]{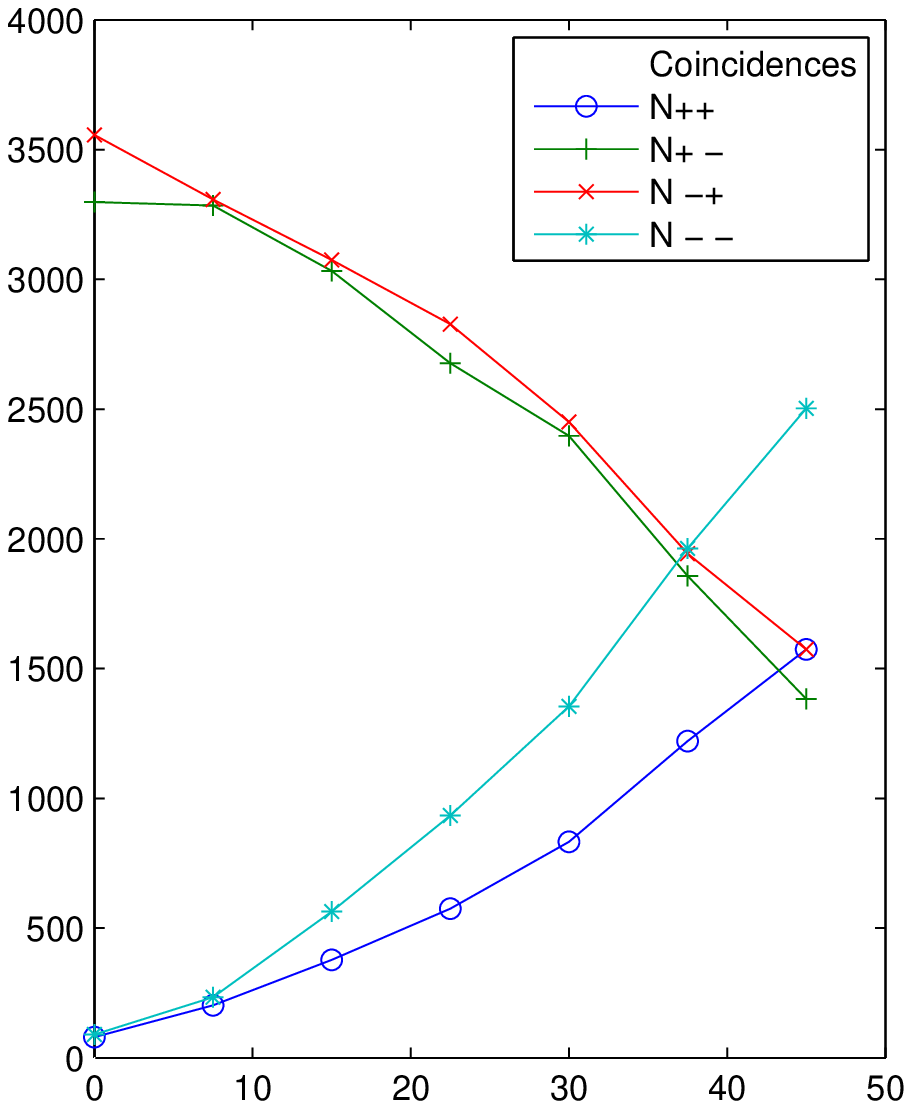}
\includegraphics[width=7cm]{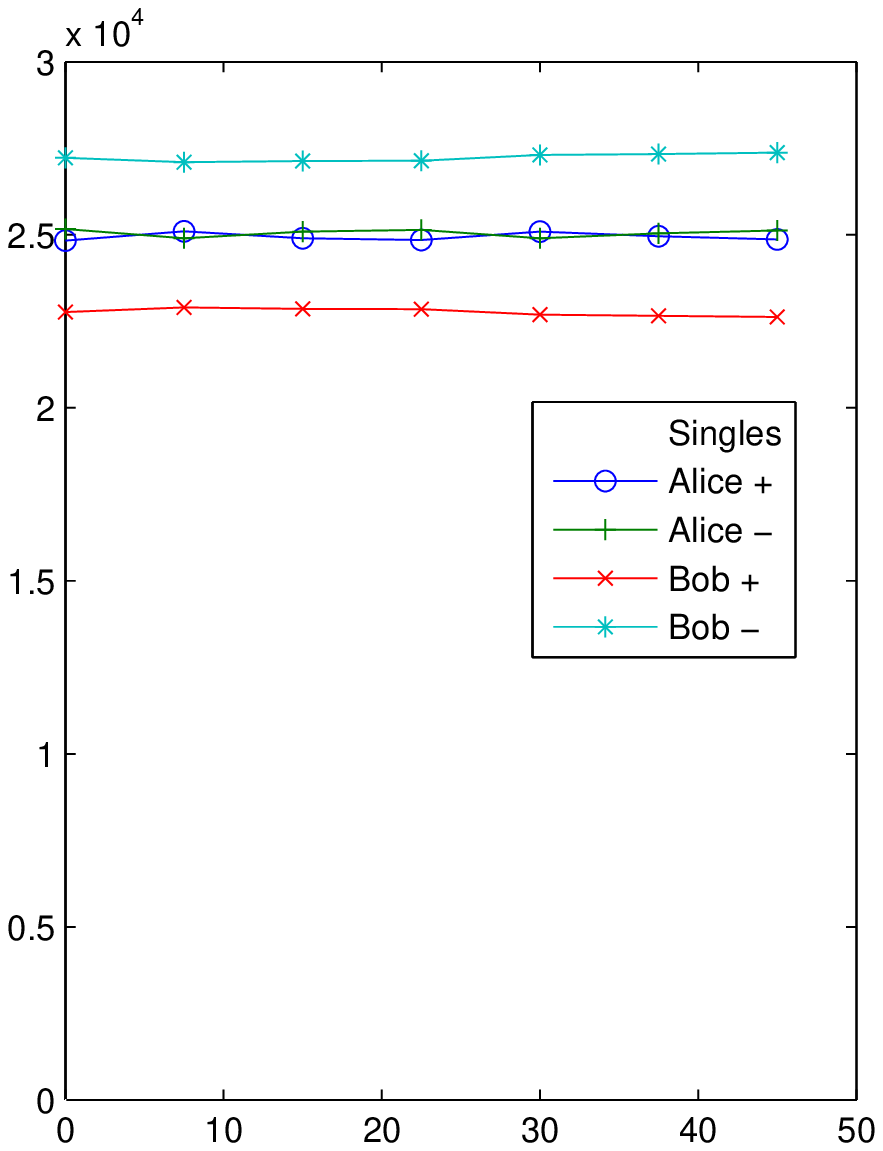}
\caption{\label{fig:coinshort} {\bf (a)} Coincidence counts as Alice
varies her measurement angle (in degrees). Note that $N_{++}$ and
$N_{--}$ differ significantly while $N_{+-}$ and $N_{-+}$ roughly
coincide. (Sourcefile: bluesine), {\bf (b)} Single Counts for Alice
and Bob as Alice varies measurement angle $\alpha$. Bob's detector
are imbalanced (he gets more clicks in the minus channel).}
\end{figure}

As can be seen in the coincidence rate figures (see figure
\ref{fig:coinshort}a), the coincidence rates exhibit minima close to
zero, and cosine-squared shape, as expected from the predictions of
Quantum Mechanics. However, just like in the long distance
experimental setup, the maxima of the four coincidence curves differ
significantly.

If we now extract the single-counts, in order to use or
normalization procedure, we can see that the efficiency of + and -
channels differ significantly for Bob (see figure
\ref{fig:coinshort}b).

\begin{figure}
\center
\includegraphics[width=8cm]{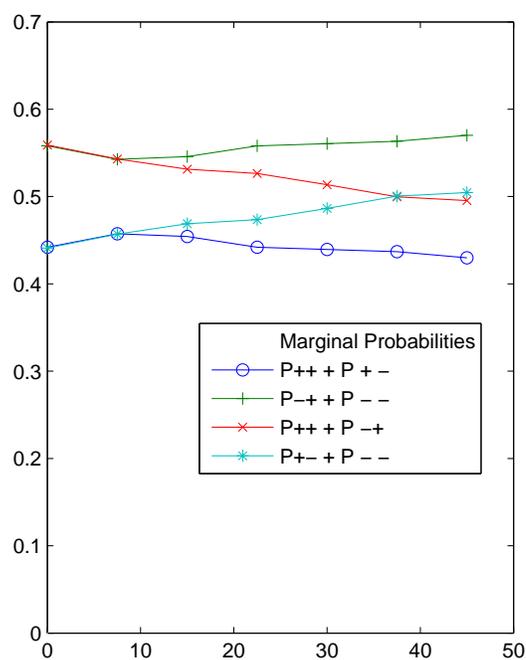}
\caption{\label{fig:margshort} Experimental result that should
coincide with the marginal probabilities, assuming Fair Sampling.}
\end{figure}

Finally, just like in the long distance setup, the \emph{four}
experimental quantities that should coincide with the marginal
probabilities (see figure \ref{fig:margshort}) vary with Alice's
measurement angle $\alpha$.

This time, the experimental quantities that should coincide with
Bob's marginal clearly depend on Alice's measurement setting
$\alpha$. It should be noted that our normalization procedure
based on the Fair Sampling Assumption reduces the magnitude of
this behavior, but without removing it completely. If the standard
normalization based only on the coincidence counts are used (by
computing quantities like $N^{++}+N^{-+}$ divided by the total sum
of coincides), the visibility of these anomalies are even greater.

\section{Conclusion}

Our result, obtained under the assumption of fair sampling, shows
that Bob's marginal probabilities clearly vary with Alice's
setting $\alpha$, in violation of the non signalling principle. It
logically means that these anomalies imply either the violation of
the fair sampling assumption, or the violation of the mo
signalling principle, or the violation of both.

It should be noted that these anomalies cannot be the result of
variations or discrepancies in the single count channel
efficiencies. The main feature of our normalization procedure is
indeed that --- once Fair Sampling has been explicitly assumed ---
the dependence on the channel efficiencies disappear from the
final expressions relating Quantum predictions to directly
measurable experimental quantities. This is an improvement on the
standard normalization procedure, which in general leads to
results depending not only the predictions for the considered
Quantum state but also on the channel efficiencies.

The no-signalling principle being a fundamental feature of Quantum
Mechanics, as well as of Local Realism, the most reasonable
interpretation of these anomalies is that the Fair Sampling
assumption should be rejected\footnote{Note that this can be
interpreted in the framework of contextual probabilities
\cite{KHR1,KHR6}. By choosing a pair of settings
$\Pi=(\alpha,\beta)$, we choose choose a selection procedure for
pairs of particles. Each selection procedure produces its own
ensemble of pairs $S_\Pi$, and the statistical properties of
$S_\Pi$ does depend on $\Pi=(\alpha,\beta)$. It therefore depends
explicitly on an experimental context $C_\Pi$ that can relate to
Bohr's reply to Einstein in 1935. It is not clear how Bohr would
have reacted to such an explicit contextuality, but since he would
probably have rejected a remote contextuality based on an explicit
nonlocality \cite{Wiseman}, we believe that our interpretation of
contextuality based on selection procedure offers a link between
Einstein's ensemble interpretation of Quantum Mechanics and Bohr's
contextual viewpoint. The point with an explicit contextuality as
this one is that Bell's derivation of his famous inequality
becomes impossible as it is derived from a fixed probability
measure $\rho$. This problem was discussed in \cite{IntProb}, p.
95, where generalizations of Bell Inequalities were obtained
within the framework of contextuality. Such generalized Bell
Inequalities are not violated by experimental data.}, as it is the
only extra assumption that we used to observe them.

In other words, we do not believe that our investigation supports
the idea of faster than light signalling, although this
possibility cannot be logically excluded. Once the Fair Sampling
assumption is rejected, there is no evidence of violation of the
no-signalling principle, and therefore no evidence whatsoever of a
possible superluminal communication. It should nevertheless be
stressed that if one prefers to maintain the Fair Sampling
assumption at all cost, the most striking result that can be
derived under this assumption is not so much the violation of Bell
Inequalities, but rather the violation of no signalling principle.

In any case, our results show that more experiments are definitely
required to understand whether this feature is unique or not in
EPR experiments.

\ack We are most grateful to Gregor Weihs from Waterloo University
(Canada) as he had the courtesy to send us two CDs containing the
raw data gathered in Innsbruck between 1997 and 1998 in various
version of the EPR setup. This work would not have been possible
otherwise. We are also very grateful to Alain Aspect who lend us a
precious copy of his PhD. We are grateful to Alain Aspect and
Gregor Weihs for stimulating discussions on the structure of
statistical data in EPR experiments.

This work was supported by the Profile Mathematical Modelling of
V\"axj\"o University and the EU-network on QP and Applications.

\appendix

\section*{Predictions of Quantum Mechanics for a non-maximally entangled
state}\label{QMIdeal}
\setcounter{section}{1}

As an example of the non-signalling principle, we can consider the
source of pairs of photons to be represented by a non-maximally
entangled state of the form:
\begin{equation}\label{source}
    |\psi\rangle=\frac{1}{\sqrt{1+p^2}}[|H\rangle_1\otimes|V\rangle_2-p|V\rangle_1\otimes|H\rangle_2].
\end{equation}
With this state, it is straightforward to obtain the joint
probability of detecting a $++$ event when polarizers 1 and 2 are
oriented at $\alpha$ and $\beta$ respectively. The result is given
in quantum mechanics by the Born rule as:
\begin{equation}\label{p++}
    P_{QT}^{++}(\alpha,\beta)=\big|(\langle+_\alpha|\otimes\langle+_\beta|)|\psi\rangle\big|^2
\end{equation}
The joint probability
$P_{QT}^{\varepsilon_1,\varepsilon_2}(\alpha,\beta)$ to get the
results $\varepsilon_1$ and $\varepsilon_2$, given the measurement
settings $\alpha$ and $\beta$, can be derived explicitly as:
\begin{eqnarray}\label{qprednme}
    P_{QT}^{++}(\alpha,\beta)&=&\frac{(p\; \sin\alpha\cos\beta- \cos\alpha\sin\beta)^2}{1+p^2}
    \nonumber
    \\
    P_{QT}^{+-}(\alpha,\beta)&=&\frac{(\cos\alpha\cos\beta+p\;\sin\alpha\sin\beta)^2}{1+p^2}
    \nonumber
    \\
    P_{QT}^{-+}(\alpha,\beta)&=&\frac{(p\;\cos\alpha\cos\beta+\sin\alpha\sin\beta)^2}{1+p^2}
    \\
    P_{QT}^{--}(\alpha,\beta)&=&\frac{(\sin\alpha\cos\beta-p\;\cos\alpha\sin\beta)^2}{1+p^2}.
    \nonumber
\end{eqnarray}
It is straightforward to show that the marginal probabilities for
this particular source state are:
\begin{eqnarray}\label{predictions}\nonumber
P^{++}_{QT}(\alpha,\beta)+P^{+-}_{QT}(\alpha,\beta)=\frac{\cos^2\alpha+p^2\sin^2
\alpha}{1+p^2}
\\
P^{-+}_{QT}(\alpha,\beta)+P^{--}_{QT}(\alpha,\beta)=\frac{p^2\cos^2\alpha+\sin^2
\alpha}{1+p^2}
\\
P^{++}_{QT}(\alpha,\beta)+P^{-+}_{QT}(\alpha,\beta)=\frac{p^2\cos^2\beta+\sin^2
\beta}{1+p^2}\nonumber
\\
P^{+-}_{QT}(\alpha,\beta)+P^{--}_{QT}(\alpha,\beta)=\frac{\cos^2\beta+p^2\sin^2
\beta}{1+p^2},\nonumber
\end{eqnarray}
which fulfil the non-signaling principle, as Alice's marginal
probabilities depend $\alpha$ alone, while Bob's marginal
probabilities depend on $\beta$ alone.

\end{document}